\documentclass[
    ,final            % use final for the camera ready runs
    ,sort&compress    % to get my citations compressed like [1-4] if they are consecutive
  ]
  {aipproc}

\layoutstyle{6x9}

\usepackage{rotating} % rotate
\begin{document}

\title{Lepton flavour violating slepton decays to test type-I and II seesaw at the LHC}


\author{Albert Villanova del Moral}{
  address={Departamento de F\'\i sica and CFTP, Instituto Superior T\'ecnico, Avenida Rovisco Pais 1, 1049-001 Lisboa, Portugal},
  ,email={albert@cftp.ist.utl.pt}
}

%\author{<author2>}{
%  address={<common address for author2 and author3>}
%}

%\author{<author3>}{
%  address={<common address for author2 and author3>}
%  ,altaddress={<author1 address>} % additional visiting address
%}

\begin{abstract}
Searches at the LHC of lepton flavour violation (LFV) in slepton decays can indirectly test both type-I and II seesaw mechanisms. Assuming universal flavour-blind boundary conditions, LFV in the neutrino sector is related to LFV in the slepton sector by means of the renormalization group equations. Ratios of LFV slepton decay rates result to be a very effective way to extract the imprint left by the neutrino sector. Some neutrino scenarios within the type-I seesaw mechanism are studied. Moreover, for both type-I and II seesaw mechanisms, a scan over the minimal supergravity parameter space is performed to estimate how large LFV slepton decay rates can be, while respecting current low-energy constraints.
\end{abstract}

\maketitle

%%%%%%%%%%%%%%%%%%%%%%%%%%%%%%%%%%%%%%%%%%%%
%% MAINMATTER
%%%%%%%%%%%%%%%%%%%%%%%%%%%%%%%%%%%%%%%%%%%%

\section{Introduction}
Neutrino experiments~\cite{Fukuda:1998mi,:2008ee,Arpesella:2008mt,Aharmim:2008kc,Adamson:2008zt} have firmly established that neutrinos have mass and their flavours mix. 
Many theoretical models have been proposed to explain current neutrino data~\cite{Schwetz:2008er}, being the seesaw mechanism~\cite{Minkowski:1977sc,GellMann:1980vs,Yanagida:1979as,Mohapatra:1979ia,Schechter:1980gr,Schechter:1981cv} one of the most popular solutions. 
%Among the theoretical attempts to explain current neutrino data~\cite{Schwetz:2008er}, the ones based in the seesaw mechanism~\cite{Minkowski:1977sc,GellMann:1980vs,Yanagida:1979as,Mohapatra:1979ia,Schechter:1980gr,Schechter:1981cv} one of the most populars solutions.
Although standard seesaw type models are not directly testable (as they require an inaccessible very high energy scale where lepton number is violated), they can be indirectly tested under certain circumstances. If universal flavour-blind boundary conditions (like mSugra) are assumed, then LFV in the neutrino sector is related to LFV in the %charged lepton and 
slepton sector~\cite{Borzumati:1986qx}. 
\section{Seesaw type-I}
\begin{figure}[htbp]
  \label{fig:DegNuR-s13}
\begin{tabular}{cc}
  \includegraphics[width=.45\textwidth]{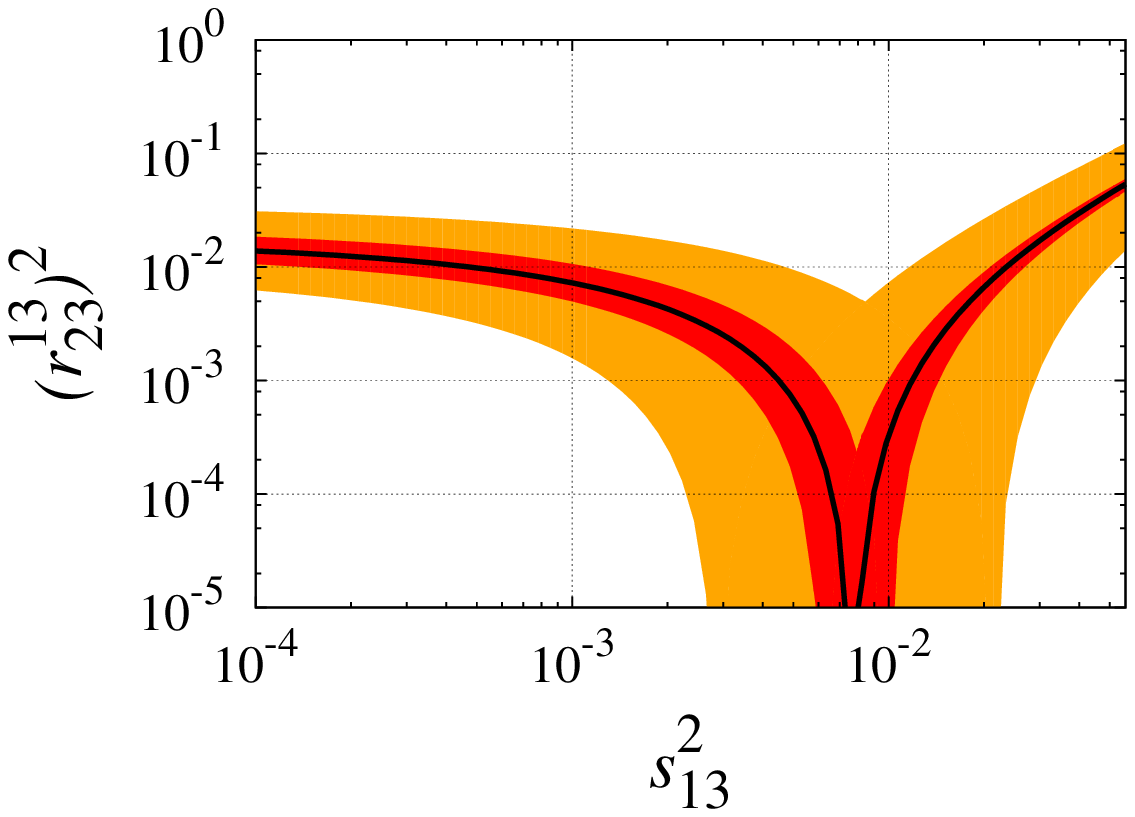} &
  \includegraphics[width=.45\textwidth]{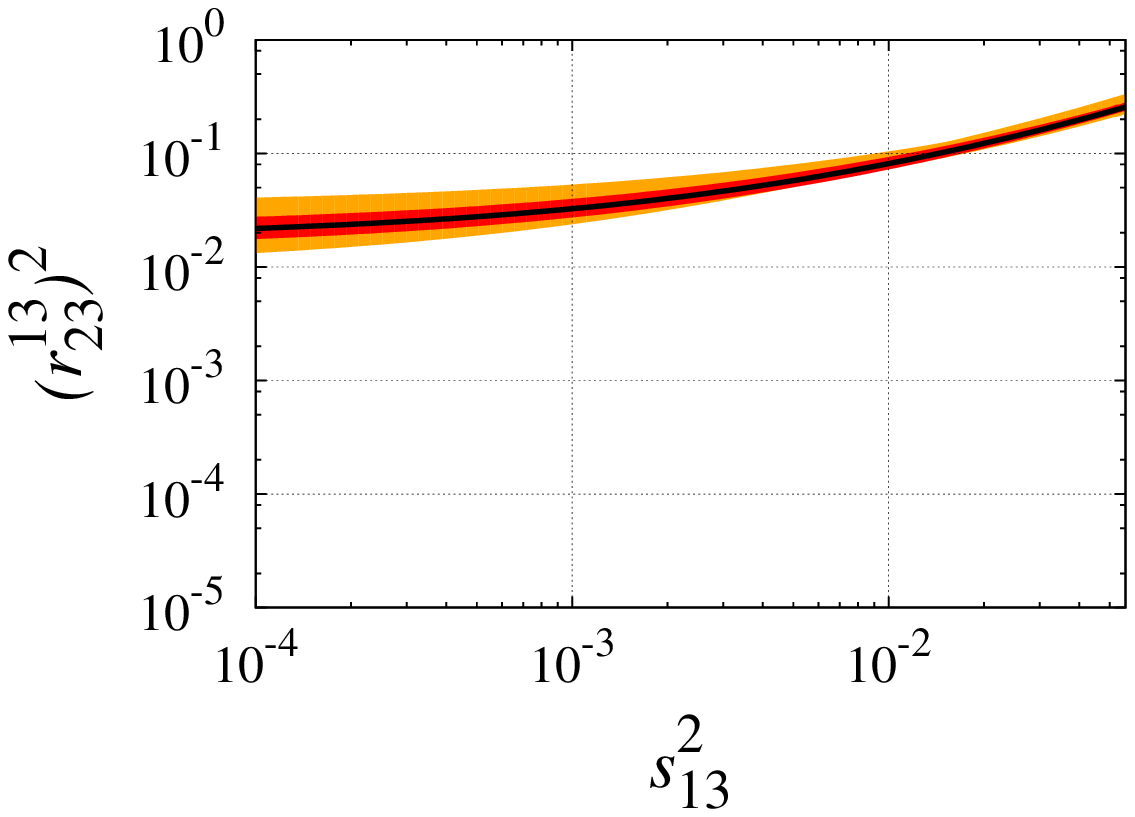}\\
  \includegraphics[width=.45\textwidth]{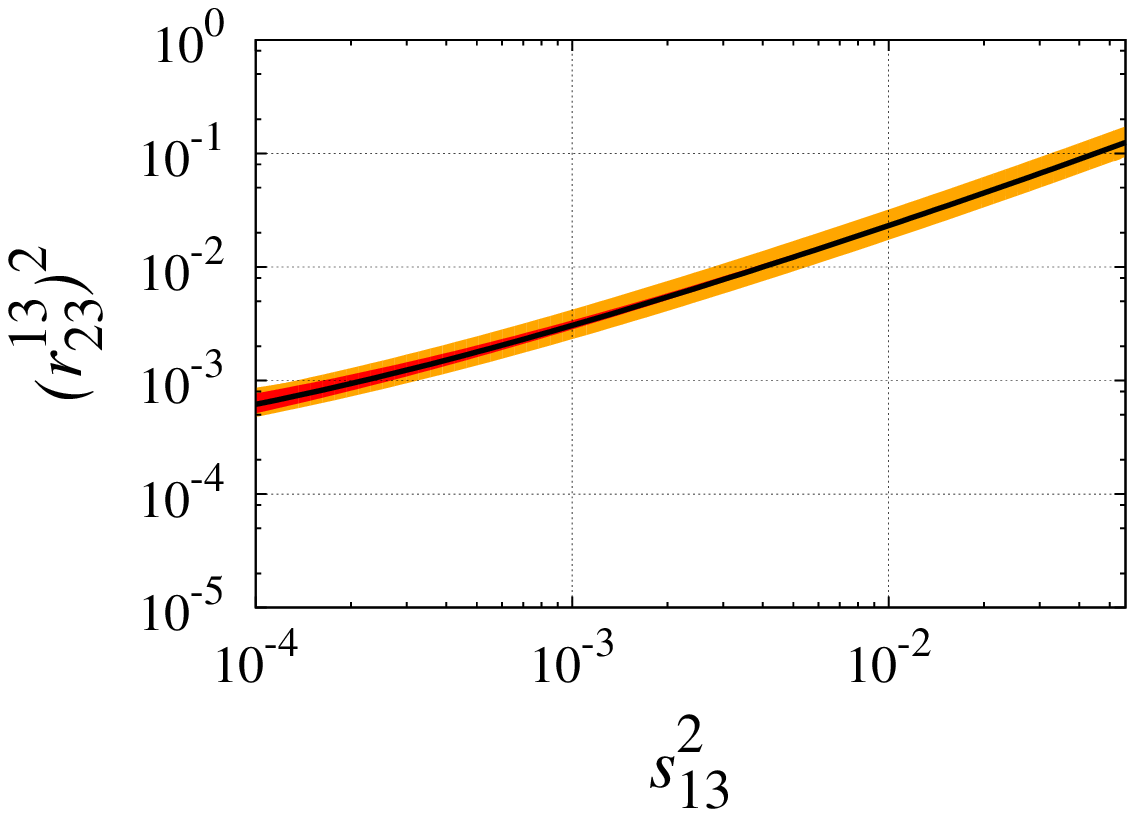}&
  \includegraphics[width=.45\textwidth]{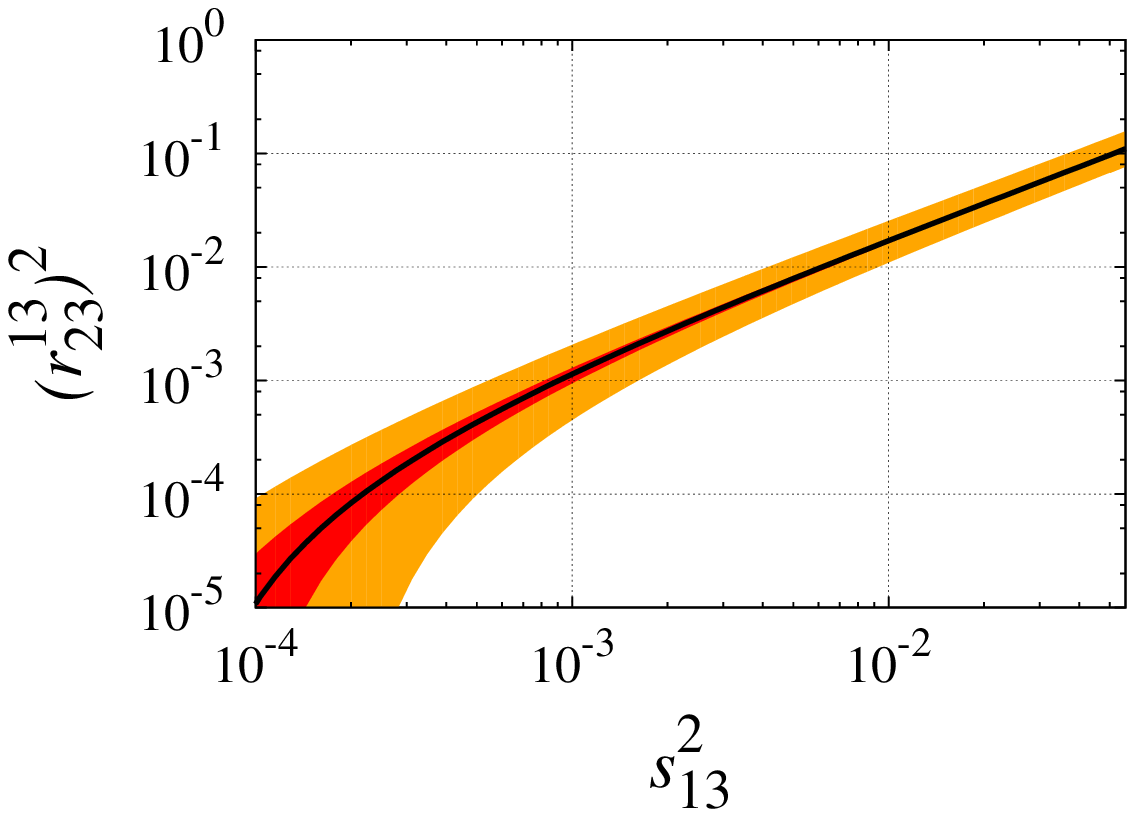}
\end{tabular}
\caption{Square ratio $(r^{13}_{23})^2$ versus $s_{13}^2$ for the case of degenerate heavy neutrinos and real $R$. 
The dark line corresponds to %the additional assumptions of 
light neutrino mass splitting fixed to their best fit point values~\cite{Schwetz:2008er} and leptonic mixing angles $\theta_{12}$ and $\theta_{23}$ fixed to their tribimaximal values~\cite{Harrison:2002er}. 
The red (dark) band corresponds to light neutrino mass splittings in their $3\sigma$ allowed range and leptonic mixing angles $\theta_{12}$ and $\theta_{23}$ fixed to their tribimaximal values. 
The orange (light) band corresponds to light neutrino mass splittings and leptonic mixing angles in their $3\sigma$ allowed range. 
%and the rest of the neutrino parameters fixed to their best fit point values. 
Each column corresponds to a different value of the Dirac phase: $\delta=0$ (first column) and $\delta=\pi$ (second column). 
Each row corresponds to a different neutrino scenario: strict normal hierarchy (first row) and strict inverse hierarchy (second row).}%, QDNH (third row) and QDIH (forth row).}
\end{figure}

Results presented in this section are based on~\cite{Hirsch:2008dy}. 
Here we study the relations between LFV in the neutrino and the slepton sectors in the framework of the $\nu$CMSSM, the Constrained Minimal Supersymmetric Standard Model with three additional singlet neutrino superfields. %, where neutrino masses are generated via the standard supersymmetric seesaw type-I mechanism. 
%We briefly introduce the simplest type-I SUSY seesaw mechanism. 
%Its particle content is the same as in the MSSM, but enlarged by three right-handed neutrino superfields ${\widehat N^c}_i$. The leptonic part of the superpotential is
One of the main inconveniences of this model is that the number of parameters at high energies is much larger than the number of observables at low energies. Nevertheless this obstacle can be circumvented by assuming certain neutrino scenarios, which fix some of the parameters. This enables to establish relations between the rest of the parameters and the observables. 
%%%%%%%%%%%%%%%%%%%%%%%%%%%%%%%%%%%%%%%%%%%%
%\subsection{Analysis}
For qualitative understanding, %we consider: the small-angle approximation (i.e., that the mixing between left-slepton flavour eigenstates is small); and the leading-log approximated solutions to the Renormalization Group Equations (RGE's), which are the responsible ones of the imprint left in the slepton sector by the neutrino sector. Under these simplifying assumtions, the left-slepton LFV decays are proportional to
the left-slepton LFV decays can be approximately expressed as
\begin{equation}
\textrm{Br}(\tilde l_i\to l_j\chi_1^0)\propto \left|(\Delta M_{\tilde L}^2)_{ij}\right|^2 \propto \left|(Y_{\nu}^{\dagger}\cdot L\cdot Y_{\nu})_{ij}\right|^2. 
\end{equation}
We can parametrize the neutrino Yukawa matrix in terms of %the maximum number of observables as~\cite{Casas:2001sr}
observables as~\cite{Casas:2001sr}
\begin{equation}\label{Ynu}
Y_{\nu} =\sqrt{2}\frac{i}{v_U}\sqrt{\hat M_R}\cdot R\cdot\sqrt{{\hat m_{\nu}}}\cdot U^{\dagger},
\end{equation}
where $\hat m_{\nu}$ and $\hat M_R$ are diagonal matrices with 
%the light neutrino mass eigenvalues $m_i$ and the heavy neutrino mass eigenvalues $M_i$, respectively; 
the light and the heavy neutrino mass eigenvalues, respectively; 
$U$ is the leptonic mixing matrix and $R$ is a complex orthogonal matrix. 
This way, the left-slepton LFV decays are related to neutrino parameters. 
%\begin{equation}
%\textrm{Br}(\tilde l_i\to l_j\chi_1^0)\propto \left|U_{i\alpha}U_{j\beta}^*\sqrt{m_{\alpha}}\sqrt{m_{\beta}}
%R_{k\alpha}^*R_{k\beta}M_k\log\left(\frac{M_X}{M_k}\right)\right|^2. 
%\end{equation}
In order to eliminate most of the dependence on the supersymmetric parameters, we work with ratios of LFV decay rates. 
Thus, for example, the ratio of stau LFV decays can be expressed in terms of the parameter $r^{13}_{23}$, 
\begin{equation}\label{eq:3}
\frac{\textrm{Br}({\tilde\tau}_2 \to  e +\chi^0_1)}
     {\textrm{Br}({\tilde\tau}_2 \to  \mu +\chi^0_1)}
 \simeq \frac{|(\Delta M_{\tilde L}^2)_{13}|^2}{|(\Delta M_{\tilde L}^2)_{23}|^2}
 \equiv \left(r^{13}_{23}\right)^2,
\end{equation}
which only depends on neutrino parameters.
As an example, %\footnote{Plots in  Fig.~\ref{fig:DegNuR-s13} have been updated with respect the ones in Ref.~\cite{Hirsch:2008dy} using more recent neutrino data~\cite{Schwetz:2008er}.}, 
Fig.~\ref{fig:DegNuR-s13} shows the expected ratio of stau LFV decays $(r^{13}_{23})^2$ as a function of the neutrino mixing angle $s_{13}^2$.%, for different neutrino scenarios and for two choices of the Dirac phase $\delta$. %Note that plots in Fig.~\ref{fig:DegNuR-s13} have been updated with respect the ones in Ref.~\cite{Hirsch:2008dy} using more recent neutrino data~\cite{Schwetz:2008er}.
%More examples can be found in~\cite{Hirsch:2008dy}.
%\begin{figure}[htbp]
%  \includegraphics[width=.5\textwidth]{plots/plot-r1323sq-m1_-_NH-IH.eps}
%  \caption{Picture to fixed height}
%\end{figure}
%
%
%Note that plots in Fig.~\ref{fig:DegNuR-s13} have been updated with respect the ones in Ref.~\cite{Hirsch:2008dy} using more recent neutrino data~\cite{Schwetz:2008er}. 
%%%%%%%%%%%%%%%%%%%%%%%%%%%%%%%%%%%%%%%%%%%%
%\subsection{Numerics}

In order to check the validity of our analytical estimated ratio of stau LFV decays, we have performed a numerical calculation with the program package \textsc{SPheno}~\cite{Porod:2003um} for the mSugra standard points SPS1a'~\cite{AguilarSaavedra:2005pw} and SPS3~\cite{Allanach:2002nj}. %, the stau LFV decays have been calculated in different neutrino scenarios. 
%Our results show that the ratio of the stau LFV decays follows very accurately the analytical estimate in the region allowed by the upper limit on Br($\mu\to e\gamma$). For more datails, see~\cite{Hirsch:2008dy}. For a similar analysis, but for seesaw type-II, see~\cite{Hirsch:2008gh}.
Our results show that the ratio of the stau LFV decays follows very accurately the analytical estimate. 
For more details, see~\cite{Hirsch:2008dy}. 
For a similar analysis, but for seesaw type-II, see~\cite{Hirsch:2008gh}.
%%%%%%%%%%%%%%%%%%%%%%%%%%%%%%%%%%%%%%%%%%%%
%%%%%%%%%%%%%%%%%%%%%%%%%%%%%%%%%%%%%%%%%%%%
\section{Scan}
Results presented in this section are based on~\cite{Esteves:2009vg}.
Although stau LFV decays have been studied for two specific SUSY benchmark points (SPS1a' and SPS3), a more general study over the mSugra parameter space is necessary. %Thus, we can identify regions of the mSugra parameter space that maximize the magnitude of ${\tilde\tau}_2$ LFV BR's. Besides this, 
For both seesaw type-I and II (details on the realization of the type-II seesaw can be found in~\cite{Rossi:2002zb}), we have estimated the maximum number of events of the opposite-sign dilepton signal $\chi^0_2\to\chi^0_1\,\mu\,\tau$, which can be searched for at the LHC\footnote{Note that a complete Monte Carlo analysis would be needed, but this is out of the scope of this work.}.
To do so, we have used program packages \textsc{SPheno}~\cite{Porod:2003um} and \textsc{Prospino}~\cite{Beenakker:1996ch,Beenakker:1997ut,Beenakker:1999xh,Spira:2002rd,Plehn:2004rp}. 
For more details, see~\cite{Esteves:2009vg}. 
%The numerical procedure consists in scanning over the $m_0$-$m_{1/2}$ plane, for fixed values of other mSUGRA parameters, using the program package \textsc{SPheno}~\cite{Porod:2003um}. %For each point in this plane, we perform a maximization of BR($\mu\to e \gamma$), so that it is as close as possible to its current experimental upper bound~\cite{Amsler:2008zzb}. %, given in equation (\ref{eq:BRMuToEGammaBound}). 
%To fit neutrino data we have followed the same iterative procedure described in section~\ref{sec:correlations-numerical}. 
%
%In order to simplify the numerical study, for light neutrinos, we have always fixed their hierarchy to be strictly normal, their mass splittings to their best fit point values~\cite{Schwetz:2008er} and their mixing to be tribimaximal~\cite{Harrison:2002er}. 
%In our analysis we have always fitted light neutrino mass splittings to their best fit point values~\cite{Schwetz:2008er}, a strictly normal hierarchical  and their mixing to be TBM.
%
%To estimate the number of events of the opposite-sign dilepton signal $\chi^0_2\to\chi^0_1\,\mu\,\tau$, we have calculated the total production cross section of $\chi^0_2$ at leading order with the package \textsc{Prospino}~\cite{Beenakker:1996ch,Beenakker:1997ut,Beenakker:1999xh,Spira:2002rd,Plehn:2004rp}.

\begin{figure}[htbp]
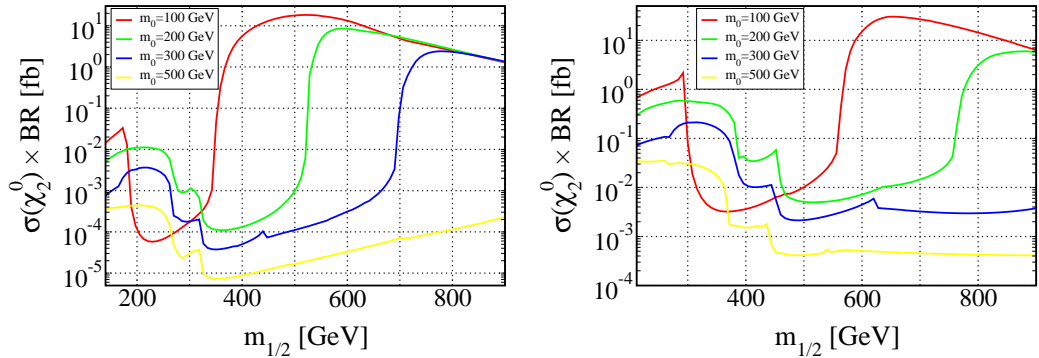

  \label{fig:ProdXBR-I-II}
  \centering
\begin{tabular}{cc}
  \includegraphics[width=0.45\textwidth]{plots/plot-sigmaLOfbBR-m12_-_4m0.eps} &
  \includegraphics[width=0.45\textwidth]{plots/plot-sigmaLOfbBR-m12_-_4m0_-_II.eps}
\end{tabular}
  \caption{Production cross section (at leading order) of $\chi^0_2$ times BR($\chi^0_2\to\chi^0_1\,\mu\,\tau$)
    %of $\chi^0_2$ going to $\mu$-$\tau$ opposite-sign dilepton signal 
    versus $m_{1/2}$ for $m_0=100$~GeV (red),
    200~GeV (green), 300~GeV (blue) and 500~GeV (yellow), in seesaw type-I (left panel) and II (right panel). We take a standard choice of
    parameters: $\mu>0$, $\tan\beta=10$ and $A_0=0$ GeV.} %, for type-I seesaw, for BR($\mu\to e \gamma) \le 1.2\cdot 10^{-11}$.} 
%  \caption{Production cross section (at leading order) of $\chi^0_2$ times BR($\chi^0_2\to\chi^0_1\,\mu\,\tau$)
%%    of $\chi^0_2$ going to $\mu$-$\tau$ opposite-sign dilepton signal 
%    versus $m_{1/2}$ for $m_0=100$~GeV (red),
%    200~GeV (green), 300~GeV (blue) and 500~GeV (yellow). We take a standard choice of
%    parameters: $\mu>0$, $\tan\beta=10$ and $A_0=0$ GeV, for type-II seesaw, for
%    BR($\mu\to e +\gamma) \le 1.2\cdot 10^{-11}$.} 
\end{figure}

Fig.~\ref{fig:ProdXBR-I-II} shows the production cross section $\sigma(\chi^0_2)$ at leading order times the branching ratio of $\chi^0_2\to\chi^0_1\,\mu\,\tau$ %going to the opposite-sign dilepton signal $\chi^0_1\,\mu\,\tau$ 
as a function of $m_{1/2}$, for different values of $m_0$, in seesaw type-I (left panel) and II (right panel). 
%
%For type-I seesaw, %we have assumed that right-handed neutrinos are degenerate and the matrix $R$ is real.
%%
%left panel in Fig.~\ref{fig:ProdXBR-I-II} shows the production cross section $\sigma(\chi^0_2)$ at leading order times the BR of $\chi^0_2$ going to the opposite-sign dilepton signal $\chi^0_1\,\mu\,\tau$ as a function of $m_{1/2}$, for different values of $m_0$. %We have fixed the rest of the mSugra parameters to a standard point defined by $\mu>0$, $\tan\beta=10$ and $A_0=0$ GeV. 
%For $m_0\sim 100$ GeV and $m_{1/2}\sim[450,\,600]$ GeV and assuming a luminosity ${\cal L} = 100 fb^{-1}$, the estimated number of events of the opposite-sign dilepton signal $\chi^0_2\to\chi^0_1\,\mu\,\tau$ can be of the order of $10^3$.
Assuming a luminosity ${\cal L} = 100 \ \textrm{fb}^{-1}$, there are regions in the parameter space where the estimated number of events of the opposite-sign dilepton signal $\chi^0_2\to\chi^0_1\,\mu\,\tau$ can be of the order of $10^3$.

%In order to simplify our numerical analysis in type-II seesaw, we have considered only the case in which $\lambda_1=2\times 10^{-2}$ and $\lambda_2=0.5$ (for details on this realization of the type-II seesaw, see~\cite{Rossi:2002zb}). We have explicitly checked that $\lambda_1$ plays no role in LFV. On the other hand, larger values of $\lambda_2$ generally increase the LFV signal, as BR($\mu\to e\,\gamma$) decreases for larger $\lambda_2$.
%
%For type-II seesaw (details on this realization of the type-II seesaw can be found in~\cite{Rossi:2002zb}), 
%right panel in Fig.~\ref{fig:ProdXBR-I-II} shows the same as the left panel but for type-II seesaw.
%Assuming a luminosity of ${\cal L} = 100 fb^{-1}$, there can be a maximum number of events of the order of $10^3$ for $m_0\sim 100$ GeV and $m_{1/2}\sim[600,\, 800]$ GeV.
%%%%%%%%%%%%%%%%%%%%%%%%%%%%%%%%%%%%%%%%%%%%%%%%
\section{Conclusion}
%Neutrino experimental data show that neutrinos are massive and mix. If its origin is the simplest supersymmetric type-I seesaw mechanism and mSugra boundary conditions hold, then LFV decays are related to neutrino parameters. In particular, we have studied the relation of the ratio of the stau LFV decays with neutrino parameters for different neutrino scenarios.
%
We have shown that the $\nu$CMSSM (SUSY seesaw type-I with mSugra boundary conditions) can be indirectly tested at the LHC by measuring the \emph{ratio} of stau LFV decay rates. 
We have performed a numerical analysis of the absolute values of stau LFV decays in both type-I and II seesaw and %we have estimated the maximum number of events that can occur at the LHC. We 
we have shown that there exist regions of the mSugra parameter space where the estimated number of events of the opposite-sign dilepton signal $\chi^0_2\to\chi^0_1\,\mu\,\tau$ can be as much as of the order of $10^3$. 
%%%%%%%%%%%%%%%%%%%%%%%%%%%%%%%%%%%%%%%%%%%%%%%%
%% BACKMATTER
%%%%%%%%%%%%%%%%%%%%%%%%%%%%%%%%%%%%%%%%%%%%%%%%

\begin{theacknowledgments}
The author wishes to thank his collaborators J.~N.~Esteves, M.~Hirsch, W.~Porod, J.~C.~Romao and J.~W.~F.~Valle.
The author is supported by {\it Funda\c c\~ao para a Ci\^encia e a Tecnologia} under the grant SFRH/BPD/30450/2006.
%The work of A.V.M. is supported by
%
This work was partially supported by FCT through the projects
CFTP-FCT UNIT 777 and CERN/FP/83503/2008, which are partially
funded through POCTI (FEDER), and by the Marie Curie RTN MRT-CT-2006-035505. 
%Work supported by the European Commission network MRTN-CT-2006-035505 and by {\it Funda\c c\~ao para a Ci\^encia e a Tecnologia} through the projects CFTP-FCT UNIT 777 and CERN/FP/83503/2008. 
\end{theacknowledgments}



\bibliographystyle{aipproc}   % if natbib is available

%\bibliography{sample}
\bibliography{avm}

\end{document}